\documentclass[showpacs,preprintnumbers,amsmath,amssymb,floatfix,nofootinbib]{revtex4-1}
\usepackage[usenames]{color}
\usepackage{epsfig}
\usepackage{amssymb}
\usepackage{multirow}
\usepackage{longtable}
\usepackage{xcolor}
\newcommand{\be}{\begin{equation}}
\newcommand{\ee}{\end{equation}}
\newcommand{\bee}{\begin{eqnarray}}
\newcommand{\eee}{\end{eqnarray}}

\definecolor{grey}{rgb}{0.9,0.9,0.9}
\definecolor{black}{rgb}{0,0,0}

\def \irbaddress{Rudjer Bo\v{s}kovi\'{c} Institute, Bijeni\v{c}ka cesta 54, P.O. Box 180, 10002 Zagreb, Croatia}
\def \untzaddress{University of Tuzla, Faculty of Science, Univerzitetska 4, 75000 Tuzla, Bosnia and Herzegovina}
\def \GWUSAIDaddress{Data Analysis Center at the Institute for Nuclear Studies, Department
of Physics, The George Washington University, Washington, D.C. 20052}

\begin{document}

\title{Pole structure from energy-dependent and single-energy  \\
fits to GWU-SAID $\pi N$ elastic scattering data}
\author{Alfred \v{S}varc}
\email{alfred.svarc@irb.hr}
\affiliation{\irbaddress}
\author{Mirza Had\v{z}imehmedovi\'{c}}
\affiliation{\untzaddress}
\author{Hedim Osmanovi\'{c}}
\affiliation{\untzaddress}
\author{Jugoslav Stahov}
\affiliation{\untzaddress}
\author{Ron L. Workman}
\affiliation{\GWUSAIDaddress}

\begin{abstract}

The pole structure of the current GWU-SAID partial-wave analysis of elastic $\pi N$ scattering and $\eta N$ production 
data is studied. Pole positions and residues are extracted from both the
energy-dependent and single-energy fits, using two different methods. For the energy-dependent fits, both contour 
integration and a Laurent+Pietarinen approach are used. In the case of 
single-energy fits, the Laurent+Pietarinen approach is used. Errors are estimated and the two sets of results 
are compared to other recent and older fits to data. 
\end{abstract}

\date{\today}

\maketitle
\section{Introduction}
The poles of partial-wave scattering amplitudes establish a 
direct link between experiment and QCD theories and models in hadron spectroscopy, and in the Review of Particle 
Physics (RPP)~\cite{PDG}, pole-related quantities are beginning to replace the older Breit-Wigner parameters listed for many reactions. 
In principle, there are two ways to extract pole parameters from experimental data: i) construct theoretical single or multi-channel models, solve them, fit the obtained 
analytic solutions to the data, and extract the pole parameters of obtained solutions through an analytic continuation of the model functions
into the complex energy plane; or ii) make a local expansion of the partial wave T-matrix in the vicinity of a pole.  
At present, poles are usually extracted using the first method 
\cite{Doering,EBAC,CMB,Zagreb,Bonn},  but considerable effort has been put into the development of alternate 
approaches, such as the  speed plot \cite{Hoehler93}, time delay \cite{Kelkar}, N/D  
method \cite{ChewMandelstam}, regularization procedure \cite{Ceci2008} or Pad\`{e} approximation \cite{Masjuan2013}.
\\ \\ \indent
In this work, we approach the problem in two ways. 
We first give the poles of the most recent George Washington University (GWU-SAID)  
$\pi N$ elastic energy-dependent (ED) amplitudes (WI08) of Refs.~\cite{GWUWI08,GWU2}  
using analytic continuation of the GWU-SAID model function into the 
complex energy plane (described in detail in Refs.~\cite{GWU,GWU1}). We then compare the outcome with the results 
of a new Laurent+Pietarinen (L+P) expansion method  
\cite{LP1,LP2,LP3}.   In addition, using the ability of the L+P method to extract pole positions from a set of 
numbers, without explicit knowledge of the analytic structure of the function which describes  
them,  we for the first time give pole positions for single-energy (SE) solutions, 
which are associated with the WI08 ED amplitudes. No reliable pole-extraction procedure, 
starting directly from single-energy solutions, has been applied to these amplitudes, 
prior to the formulation of this method. 
\\ \\ \indent
Complete sets of pole parameters have not been published for the most recent 
GWU-SAID partial-wave analyses solution WI08~\cite{GWUWI08,GWU2}. 
The GWU-SAID results, listed in the
RPP, were obtained in 2006 and based on the SP06 ED
solution \cite{GWU,GWU1,GWU2}. One goal of this study is thus to update these pole parameters, from both the ED and 
SE amplitudes, associated with
the current fit, WI08~\cite{GWUWI08,GWU2}. As the results from the earlier SP06 fit were quoted without an error
estimate, we obtain error estimates here in order to facilitate a
comparison with other determinations listed in the RPP.
\\ \\ \indent
We emphasize that the use of both the analytic continuation and L+P methods for the ED amplitudes provides 
a consistency check and an estimate of errors~\cite{LP2}, where the two methods can be compared. The SE amplitudes are 
given at discrete energies and are not
smooth functions of energy; the contour integral is not a possibility and we take only the L+P results. As we describe
below, the SE 
amplitudes differ from ED values both in having an uncertainly tied more directly to the data, and having additional
structure, which has been used in the past to check for systematic problems in the ED fit. This additional structure
can lead to further poles being required for a best L+P fit. In the following, we compare these
added structures to previous findings in the literature.

\section{Formalism}
\subsection{The GWU-SAID ED and SE amplitudes}
In the GWU-SAID approach, $\pi N$ elastic scattering, and $\eta N$ production data covering a limited energy range, are fitted
using
a formalism based on a Chew-Mandelstam K-matrix approach, having the ED cuts at the $\pi N$ threshold plus the $\pi \Delta$ and $\rho N$ branch points. In this way, the amplitudes have unitarity constraints, 
though no inelastic 
channels beyond $\eta N$ have been fitted.
\\ \\ \indent
This choice determines how energy-dependence is introduced. The scattering into different channels is represented by a matrix $\bar{T}$, parametrized in terms of $\bar{K}$ , a (4 x 4) real symmetric K-matrix 
for each partial wave:
\begin{eqnarray}
\bar{T} &=&\bar{K}(1-C \bar{K})^{-1}, 
\end{eqnarray}
C being a diagonal Chew-Mandelstam function, with Im $C_i=\rho_i$ giving the phase-space function for the i-th channel
($\pi N$,  $\eta N$, $\pi \Delta$, and $\rho N$). 
\\ \\ \indent
In order to control the behavior of each T-matrix element near threshold, the K-matrix elements are expanded as polynomials in the energy variable $Z=(W-W_{th})$, where $W$ and $W_{th}$ are the 
center-of-mass ($\sqrt{s}$) and threshold energies, respectively. Expanding in factors of Z allows the fixing
of scattering lengths through the value of the leading term. 
\\ \\ \indent
In general, a fit of the K-matrix elements, expanded in terms of an energy variable, may not result in a form satisfying all of the requirements imposed by analyticity. To remedy this problem, the analycity 
requirement is imposed at fixed four-momentum transfer $t$ by introducing a complete set of fixed-$t$ dispersion relations (DRs), which are handled iteratively with the data fitting, as has been described in 
Ref. \cite{GWU1}. The DRs contain subtraction constants which should be independent of energy (but which can be functions of four-momentum
transfer). After each data-fitting iteration, these constants are calculated as a function of energy. $\chi^2$ deviations from the average, at a series of energies, are then calculated and included as 
pseudo-data. The partial-wave amplitudes and the real parts of the DR invariant amplitudes are then adjusted to minimize the $\chi^2$ from the sum of data and pseudodata. Compatibility with the DR 
constraints can be controlled through the errors assigned to pseudo-data in
the fit.
\\ \\ \indent
As the model is constrained by dispersion relations, the result also obeys the conditions of  analyticity. All extracted poles are on the first unphysical Riemann sheet (second Riemann sheet) with the 
exception of the Roper resonance which has two nearby poles: one pole near the $\pi \Delta$ branch point on the second Riemann sheet, but the second pole can be revealed by rotating the $\pi \Delta$  branch 
cut, and it appears on the third Riemann sheet.
\\ \\ \indent
Since the form of the K-matrix elements used in the energy-dependent fits is simple, it can be analytically continued into the complex energy plane, and it is straightforward to locate the complex energy 
positions for the poles and zeros which influence the on-shell behavior of the amplitudes.  Complex-plane contour plots of ln($|T|^2$) are generated, a starting energy near the pole/zero is picked, and the 
Newton-Raphson algorithm is used to converge on the structure. Results for the pole positions (and residues) are given in Tables \ref{tab:pole1}-\ref{tab:pole2}.

The SE solutions are generated, primarily to search for systematic deviations from the ED result, by fitting data within narrow energy bins, starting from
the ED solution. Soft constraints are added to keep these fits close to their initial ED values, though these added constraints generally have little effect on
the fit result. Energy dependence over the bin is taken to be linear, based on the starting ED value. Here, also, no dispersion-relation constraints are applied and
the energy-to-energy variation of individual partial-wave amplitudes need not be smooth. However, the fit to data is improved and one can check for any 
energy-dependent structures not contained in the underlying ED parameterization. In earlier investigations of the $P_{11}$ partial wave, it has been noticed
that the energy-dependence of GW-SAID SE amplitudes is strikingly similar to that found in the KH80 solution - an analysis 
finding resonances beyond the single Roper resonance
reported in an analysis of the WI08 ED amplitudes. Below, we expand this study and indicate further resonance signals beyond those reported for the ED SP06 and
WI08 fits.

\subsection{Laurent + Pietarinen method}

In Ref.~\cite{LP1}  we have presented a new approach to quantifying pole parameters of single-channel processes, based on replacing the global principle of analytic continuation applied to 
model functions with a local Laurent expansion of partial-wave T-matrices, in the vicinity of the real axis.  Here we specifically apply the method to ED and SE solutions of the GW-SAID group. 
The main idea of the approach is 
not to construct and solve an elaborate physics model globally, over the full 
complex-energy plane on all Riemann sheets, but instead to make use of well-known partial-wave analytic structure, and 
obtain analytic, quickly converging expansions representing partial-wave T matrices only locally, in the vicinity of the real axis and poles on the first unphysical sheet. 
We separate the divergent and regular parts of a partial-wave scattering amplitude using a Laurent (Mittag-Lefler) decomposition, 
and expand the regular part using a rapidly convergent Pietarinen series in order to obtain the most 
general function with well-defined branch points and branch cuts (which we choose) to fit the input amplitudes. 
The number of expansion terms is defined by the nature of the problem, 
and obtained by fitting the chosen ED or SE partial-wave. In this way, we represent the partial-wave scattering amplitude with the 
simplest analytic function which contains a minimal number of poles and has reasonable analytic properties, but is defined only over a limited region of the complex energy plane.  
By fitting, we obtain a certain number of poles and branch-points 
which effectively represent all inelastic channels, as determined by the input. 
The number and position of branch points is, of course, the approximation, and their choice represents the model dependence of L+P method. We use three Pietarinen functions, 
one with a branch point in the unphysical region to represent all left-hand cuts, 
and two with branch points in the physical region to represent all inelastic channels. The second branch point is usually placed at 
the $\pi N$ elastic threshold, and the third one is placed either at the dominant inelastic channel opening, or allowed to be free.   In this way we were able to introduce an error analysis into the method. 
Each choice of branch points generates a different set of pole parameters (for details see Refs.~\cite{LP1,LP2,LP3}).  The discrepancy criteria are defined  using a discrepancy parameter $D_{dp}$ for ED 
solutions, and a $\chi^2_{dp}$ quantity for SE solutions.  Unitarity is imposed in the elastic region via a penalty function. With this method, 
we are only able to extract poles are on the first unphysical Riemann sheet.

\subsubsection{Laurent and Mittag-Leffler expansion}

We expand the full partial-wave T-matrix around singularities in the second Riemann sheet using a power series expansion. Instead of using the Laurent expansion of a function near a pole, we generalize to 
the multi-pole case using the Mittag-Leffler theorem  \cite{LP1,Mittag-Leffler}  which expresses a function in  
terms of its first $k$ poles 
and an entire function:

\begin{eqnarray}
\label{eq:Laurent}
T(W) &=& \sum _{i=1}^{k} \frac{a_{-1}^{(i)}}{W -W_i}+B^{L}(W);    \, \, \, \, a_{-1}^{(i)}, W _i, W \in  \mathbb{C}.
\end{eqnarray}

Here, $W$ is c.m. energy, $a_{-1}^{(i)}$ and $W_ i$ are residue and pole position for the $i$-th pole,  and $B^{L}(W )$ is a function regular in all $W  \neq W _i$. 
It is important to note that this expansion is not a representation of the unknown function $T(W)$ in the full complex energy plane, but is restricted to the part of the complex energy plane  
where the expansion converges. 
The Laurent series converges on the open annulus around the expansion point, and if we chose to expand around the pole the outer radius of the annulus extends to the position of the next singularity (such as 
a nearby pole ), and inner radius collapses to a point.  
Hence, our Laurent expansion converges on a sum of circles located at the poles, and this part of the complex energy plane in principle includes the real axes. 
By fitting the expansion (\ref{eq:Laurent}) to the experimental data on the real axis, we in principle obtain the exact values of $S$-matrix poles.

\subsubsection{Pietarinen series}

A specific type of conformal mapping technique was proposed and introduced by Ciulli \cite{Ciulli,Ciulli1} and Pietarinen \cite{Pietarinen}, and used in the Karlsruhe-Helsinki partial wave  
analysis \cite{Hoehler84} as an efficient expansion of full
invariant amplitudes.  
It was later used by a  number of authors to solve problems in scattering and field theory \cite{ConMap-Use}, but not applied to the pole search prior to our first study~\cite{LP1}. 
A more detailed discussion of the use of conformal mapping and this method can be found in Refs.\cite{LP1,LP2,LP3}.
\\ \\ \noindent
At this point let us explain what this conformal mapping  (named Pietarinen expansion in Ref. \cite{Hoehler84}) actually means in our case.
\\ \\ \indent
If $F(W)$ is a general unknown analytic function with a cut starting at $W=x_P$, it can be represented as a power series of simple ``Pietarinen functions" all having a branch point at $W=x_P$:
\begin{eqnarray}
\label{eq:Pietarinen}
F(W ) &=& \sum_{n=0}^{N}c_n\, X(W )^n, \, \, \, \, \, \, \, \, \, \, W  \in  \mathbb{C}   \nonumber \\
X(W )&=& \frac{\alpha-\sqrt{x_P-W }}{\alpha+\sqrt{x_P-W }}, \, \, \, \, \, c_n, x_P, \alpha \in  \mathbb{R},
\end{eqnarray}
with $\alpha$ and $c_n$ acting as a tuning parameter and coefficients of the Pietarinen function $X(W)$ respectively.
\\ \\ \indent
The essence of the approach is that $(X(W )^n, \,  {n=1,\ldots, N)}$ forms a complete set of functions defined on the unit circle in the complex energy plane with a  branch cut starting at $W= x_P$, so the 
initially unknown form of any analytic function $F(W)$ having a branch point at  $W= x_P$ can be obtained by expanding it in a rapidly converging power series of simpler $X(W)$ with real coefficients, up to 
the order $N$, which are determined by fitting the input data. The unknown form of $F(W)$ should not be guessed, but its closest approximation in the form of a well defined power series is obtained only by 
fitting. 
In the calculation of  Ref.~\cite{Pietarinen}, as many as 50 terms were used; in the present analysis, covering a narrower energy range, fewer terms are required.

\subsubsection{Application of Pietarinen series to scattering theory}

The analytic structure of each partial wave is well known. 
Every partial wave contains poles which parameterize resonant contributions, cuts in the physical region starting at thresholds of elastic and all possible inelastic channels, plus
$t$-channel, $u$-channel and nucleon exchange contributions quantified with corresponding negative energy cuts. 
However, the explicit analytic form of each cut contribution is not known. 
Instead of guessing the exact analytic form of all of these, we use one Pietarinen series to represent each cut, and the number of terms in the Pietarinen series is determined by the quality of  
fit to the input data. 
In principle we have one Pietarinen series per cut; branch points $x_P, x_Q ...$ are known from physics, and coefficients are determined by fitting the input data.  
In practice, we have too many cuts (especially in the negative energy range), thus we reduce their number by dividing them into two categories: 
all negative energy cuts are approximated with only one, effective negative energy cut represented by one (Pietarinen) series (we denote its branch point as $x_P$), while each physical cut is  
represented by a separate
 series with branch points determined by the physics of the process ($x_Q,x_R...$).
\\ \\ \indent
In summary, the set of equations which define the Laurent expansion + Pietarinen series method (L+P method) is:

\begin{eqnarray}
\label{eq:Laurent-Pietarinen}
T(W ) &=& \sum _{i=1}^{k} \frac{a_{-1}^{(i)}}{W-W_i}+ B^{L}(W) 
\nonumber \\
B^{L}(W)&=& \sum _{n=0}^{M}c_n\, X(W )^n  +  \sum _{n=0}^{N}d_n\, Y(W )^n +  \sum _{n=0}^{P}e_n\, Z(W )^n  + \cdots    
\nonumber  \\
X(W )&=& \frac{\alpha-\sqrt{x_P-W}}{\alpha+\sqrt{x_P - W }}; \, \, \, \, \,   Y(W ) =  \frac{\beta-\sqrt{x_Q-W }}{\beta+\sqrt{x_Q-W }};  \, \, \, \, \,   Z(W ) =  \frac{\gamma-\sqrt{x_R-W  
}}{\gamma+\sqrt{x_R-W }} + \cdots  
\nonumber \\
&& a_{-1}^{(i)}, W _i, W   \in   \mathbb{C} 
\nonumber \\
&& c_n, d_n, e_n \alpha, \beta, \gamma ... \in  \mathbb{R}  \, \, 
{\rm and} \, \, x_P, x_Q, x_R  \in \mathbb{R} \, \,  {\rm  or} \, \, \mathbb{C}   
\nonumber \\
&& {\rm and} \, \, \, k, M, N, P  ... \in  \mathbb{N}.
\end{eqnarray}

As our input data are on the real axes, the fit is performed only on this dense subset of the complex energy plane. 
All Pietarinen parameters in equations (\ref{eq:Laurent-Pietarinen}) are determined by the fit.
\\ \\ \indent
We observe that the class of input functions which may be analyzed with this method is quite wide.
One may either fit partial wave amplitudes obtained from theoretical models, or possibly experimental data directly. 
In either case, the $T$-matrix is represented by this set of equations (\ref{eq:Laurent-Pietarinen}), and minimization is
 usually carried out in terms of $\chi^2$.
\\ \\ \indent
 In practice we use only three Pietarinen functions (one with a branch point in the unphysical region to represent all left-hand cuts, and two with  branch points in the physical region to represent the 
dominant inelastic channels), combined with the minimal number of poles. And this is the main approximation of the model used in performing the error analysis.

\subsubsection{Real and complex branch points}
While the fit strategy based on Eqs. (\ref{eq:Laurent-Pietarinen}) implies the use of purely real branch points, we know that complex branch points in the complex energy plane also exist. This feature can be 
simply seen by starting from three-body unitarity conditions, see Ref.~\cite{Gribov}. However, this possibility is also covered by our model because branch points $x_P$, $x_Q$, $x_R$ ....  in the Pietarinen 
expansion (\ref{eq:Laurent-Pietarinen})  can be either real or complex, but each of the possibilities  describe different physical situation. 
\\ \\ \indent
If the branch points $x_P$, $x_Q$, $x_R$ .... are real numbers, this means that our initial and final state contributions are defined by stable initial and final state particles. 
Then, all our background contributions are described by stable particles as well.
\\ \\ \indent
However, from experience we know that this as a general principle is not true: a three body final state is always created provided that the energy balance allows for it. In addition, when three body final 
states are formed,  they typically consist of one stable particle (nucleon or pion), and several combinations of two-body resonant substates like $\sigma$, $\rho$, $\Delta$..., and it is well known that such 
a situation is described by complex branch points \cite{Gribov,Ceci2011}. So, when we choose one of our branch points to be complex, we describe, for example, 
a three-body final-state consisting of one stable particle and one two-body resonant substate. 
\\ \\ \indent
If we suspect that three body final state might contribute significantly to the analyzed process, we choose a model where the first two branch points $x_P$ and  $x_Q$ are always real, but the third branch 
point $x_R$ can be either real (two-body final states) or complex (three-body  
final state with a resonance in a two-body subsystem).  
\\ \\ \indent
However, at this point we should stress that the single-channel character of the method forbids us to distinguish between the two. Specifically, purely real or a combination of real and complex branch points 
might equally well reproduce the single-channel data input we fit, and only the knowledge of other channels may resolve this dilemma. 
\\ \\ \indent
If only single-channel information is available, we hence have two alternatives: either we obtain a good fit with an extra resonance and stable initial and final state particles (real branch points),  or we 
obtain a good fit with one resonance less, and a complex branch point.  Data from only a single-channel do not distinguish between the two.  This effect has been already examined, elaborated and discussed in 
the case of J\"{u}lich model. There, a more detailed treatment shows how the $\rho$N complex branch point interferes and intermixes  
with the $P_{11}$(1710) 1/2$^+$, as shown in Refs.1~\cite{Ceci2011,Roenchen2013}.
\\ \\ \indent
Issues connected with the importance of inelastic channels, and two-body resonant sub-states in three-body final states, have already been recognized in ref. \cite{Hoehler93} (paragraphs 4.2 and  
4.3), however, at that time, a formalism to follow and quantify these effects did not exist, so no estimates were given.  
\\ \\ \indent
Use of the L+P formalism, with complex branch points, enables us to study these effects in detail.
We cannot distinguish whether the new resonant state manifests itself as a new isobar resonance with stable initial and final states (real branch points), or as a resonance in a two-body  
subchannel of three-body final state (complex branch point). 
For that, we need the data from extra channels, and experiments giving information on ratios of 2-body/3-body cross sections at the same energies are badly needed. 
The advantage of the Pietarinen expansion method is that it can be extended directly to complex branch points, and we use it to search partial waves where ambiguities may exist.

\subsubsection{The fitting procedure}

We use three Pietarinen functions (one with a branch point in the unphysical region to represent all left-hand cuts, and two with  branch points in the physical region to represent the dominant  
inelastic channels), combined with the minimal number of poles. 
We also allow the possibility  that one of the branch points becomes a complex number allowing all three-body final states to be effectively taken into account. 
We generally start with 5 Pietarinen terms per decomposition, and the anticipated number of poles. We minimize the discrepancy parameter $D_{dp}$ defined as:
\begin{eqnarray}
\label{discrepancy}
D_{dp} & =& \frac{1}{2 \, N_{data}} \, \, \sum_{i=1}^{N_{data}} \left[ \left( \frac{{\rm Re} \, T_i^{fit}-{\rm Re} \, T_i}{Err_i^{\rm Re}}  \right)^2  
 +   \left( \frac{{\rm Im} \, T_i^{fit}-{\rm Im} \, T_i}{Err_i^{\rm Im}} \right)^2 \right] \nonumber \\ 
 & + & \sum_{j=1}^{3}\lambda^j \, \chi_{Pen}^j +    \,  \Upsilon \, \, \sum_{j=1}^{N_{pts}^{el}}(1-S(W_j)S(W_j)^\dagger)^2 ,  
\end{eqnarray}
where $N_{data}$ is the number of energies.
 Amplitude errors are introduced differently for ED and SE solution.
 \\ \\ \noindent
  For WI08 ED solution errors are introduced as:
\begin{eqnarray}  \label{def:Err}
Err_i^{\rm Re} & = &  0.05 \, \, \frac{ \sum_{k=1}^{N_{data}} |{\rm Re} T_k^{\rm WI08}|} {N_{data}} + 0.05 \, \, |{\rm Re} T_i^{\rm WI08}| 
\nonumber \\
Err_i^{\rm Im} & = &  0.05 \, \, \frac{ \sum_{k=1}^{N_{data}} |{\rm Im} T_k^{\rm WI08}|} {N_{data}} + 0.05 \, \, |{\rm Im} T_i^{\rm WI08}| . 
\end{eqnarray}
When errors of the input numbers are not given, and one wants to make a minimization, errors have to be estimated. 
There are two simple ways to do this: either assign a constant error to each data point, or introducing an energy-dependent error as a percentage of the given value. 
Both definitions have drawbacks. 
For the first recipe only high-valued points are favored, while in the latter case low-valued points tend to be almost exactly reproduced. 
We find neither satisfactory, so we follow prescriptions used by GWU and Mainz groups, and use a combined error which consists of a sum of constant and energy dependent errors.
\\ \\ \noindent
The errors for corresponding SE solutions are taken directly from analyses.
\\ \\ \indent
The second term $\chi_{Pen}^j =  \sum _{k=1}^{N} (c_k^j)^2 \, k^3$ in the discrepancy function  $D_{dp}$ is the Pietarinen penalty function (see Ref.~\cite{Pietarinen}) which guarantees the soft cut-off of 
higher order terms in the Pietarinen expansion. The third term is the unitarity constraint. Parameters $\lambda^j$ and $\Upsilon$ are penalty function adjusting parameters which serve to bring into correct 
proportion contributions from penalty functions and contributions originating from the data itself. They are determined empirically, prior the fit, independently for each penalty function.
\\ \\ \indent
This quantity is minimized using MINUIT and the quality of the fit is visually inspected by comparing the fitting function to the data. 
If the fit is unsatisfactory (discrepancy parameters are too high, or the fit visually  does not reproduce the fitted data), the number of Pietarinen terms is increased, and if this is insufficient, the  
number of poles is increased by one.  The fit is then repeated, and the quality of the fit is re-estimated.  This procedure is continued until we reach a satisfactory fit.
\\ \\ \indent
Pole positions, residues, and Pietarinen coefficients $\alpha$, $\beta$, $\gamma$, $c_i$, $d_i$ and $e_i$ are 
our fitting parameters. 
However, in the strict spirit of the method, Pietarinen branch points $x_P$, $x_Q$ and $x_R$ should not be fitting parameters; 
each known cut should be represented by its own Pietarinen series, fixed to known physical branch points. 
While this would be ideal, in practice the application is somewhat different.  
We can never include all physical cuts from the multi-channel process. 
Instead, we represent them by a smaller subset. 
So, in our  method, Pietarinen branch points $x_P$, $x_Q$ and $x_R$ are not generally constants;  we have explored the effect of allowing them to vary as fitting parameters. 
In the following, we shall demonstrate that when searched, the branch points in the physical region still naturally converge towards branch points which belong to channels which dominate a  
particular partial wave, but may not actually correspond to them exactly.
The proximity of the fit results to  exact physical branch points describes the goodness of fit; it tells us how  well certain combinations of thresholds 
indeed approximates a partial wave.
Together with the choice of degree for the Pietarinen polynomial, this represents the model dependence of our method. 
We do not claim that our method is entirely model independent. However, the method chooses the simplest function with the given analytic properties which fit the data, and increases the  
complexity of the function only when the data require it.  

\subsubsection{Two body unitarity}
As the partial wave T matrices $T(W)$  given by equations (\ref{eq:Laurent-Pietarinen}) do not fulfill the unitarity condition manifestly, we are forced to impose elastic unitarity below the first inelastic 
threshold numerically by introducing a penalty function $\,  \Upsilon \, \, \sum_{j=1}^{N_{pts}^{el}}(1-S(W_j)S(W_j)^\dagger)^2$ into the  discrepancy parameter $D_{dp}$ of the fit defined in Eq. 
(\ref{discrepancy}). Above the first inelastic threshold this term is set to zero. 

\subsubsection{Radius of convergence and reflection principle}

As we have noted, the Mittag-Leffler expansion is not a representation of the unknown function $T(W)$ in the full complex energy plane; it is restricted to the part of the complex energy plane  
where the expansion converges. The Laurent (Mittag-Leffler) series converge on the open annulus around each point, and if we choose poles as expansion points, the open annulus collapses to a circle around 
each pole. Thus, our Laurent expansion converges on a sum of circles located at the poles, and this part of the complex energy plane in principle includes the real axes.
By fitting the expansion (\ref{eq:Laurent}) to the experimental data on the real axis, this in principle gives exact values of $S$-matrix poles.
\\ \\ \noindent
This has, however, a direct consequence on the reflection principle considerations. 
\\ \\ \indent
We know that reflection principle states that when we have a holomorphic function in part of the complex energy plane, and it is continuous and real valued on the real axis (unitarity), one can write an 
analytic continuation for the whole $\mathbb{C}$ plane such that $f(z^*) = f^*(z)$. One may consider whether 
our L+P expansion obeys this principle. However, our expansion is presently defined and valid only over a part of the complex energy plane (sum of circles on the second Riemann sheet located at poles, and 
extending to the nearest singularity), so information on the function $T^{L+P}$ at the reflection point $z^*$ is unavailable to us. One could make an analytic continuation of the function to the whole second 
Riemann sheet from $z$ to the reflection point $z^*$  through the subthreshold region where imaginary part of T matrix vanishes, but we have not attempted this as our aim is only to obtain pole parameters, 
and not to give the T matrices everywhere in  $\mathbb{C}$.
\\ \\ 
\noindent
\subsubsection{Error analysis}
In our principal paper \cite{LP1}, we have tested the validity of the model on a number of well-known $\pi$N amplitudes, and concluded that the method is very robust. 
That paper, however, did not present an error analysis. 
This has been carried out in Refs. \cite{LP2,LP3}, and here for completeness we repeat the essential details.
\\ \\ \noindent
In the L+P method we have statistical and systematic uncertainties: 1. statistical; and 2. systematic.
\\ \\
\noindent
\textit{1. Statistical uncertainty} 
\\ \\
Statistical uncertainties are simply taken  from MINUIT, which is used for minimization. This is shown separately in all tables as the first term. 
\\ \\ \noindent
\textit{2. Systematic uncertainty}
\\ \\
Systematic uncertainty is the error of the method itself, and requires a more detailed explanations. 
\\ \\
Our 
Laurent decomposition contains only two branch points in the physical region, and this is not expected to be sufficient
in a realistic case. 
Any realistic analytic function in principle contains more than two branch points, but we have approximated it in our model by using only two. 
 \\ \\ \noindent
Thus, we use the following procedure to define systematic uncertainties:

\begin{itemize}
               
\item [i)]We always allow the first (unphysical) branch point $x_P$  to be free in the uphysical region because we have no control over background contributions;
\item [ii)] We always keep the first physical branch point $x_Q$ fixed at $x_Q=1077$ MeV (the $\pi$N threshold).
\item [iii)] The error analysis is done by varying the remaining physical branch point $x_R$ in two ways:
			       
 \begin{enumerate}
					     
\item We fix the third branch point $x_R$ to the threshold of the dominant inelastic channel for the chosen partial wave (e.g. the $\eta$ threshold for the S-wave) if only one inelastic channel is  
important, or in case of several equally important inelastic processes we perform several runs with  the $x_R$  branch point fixed to each threshold in succession.
\item  We release the third branch point  $x_R$  allowing MINUIT to find an effective branch point representing all inelastic channels. 
If only one channel is dominant, the result of the fit will be close to the dominant inelastic channel.
                    
\end{enumerate}
					
\item [iv)] We average results of the fit, and obtain the standard deviation.
\end{itemize}

The choice of all values for the branch point $x_R$ is given in  the Appendix,  Table \ref{tab:pole1LPpar} for ED solutions and Table \ref{tab:pole2LPpar} for SE results. The quality of our fits is  measured 
by the discrepancy parameter $D_{dp}$ defined in Eqs.(\ref{def:Err}), and is given in the tables as well.

\section{Results and Discussion}
\subsection{Real branch points}
\begin{figure*}[!p]
\includegraphics[width=0.88\textwidth]{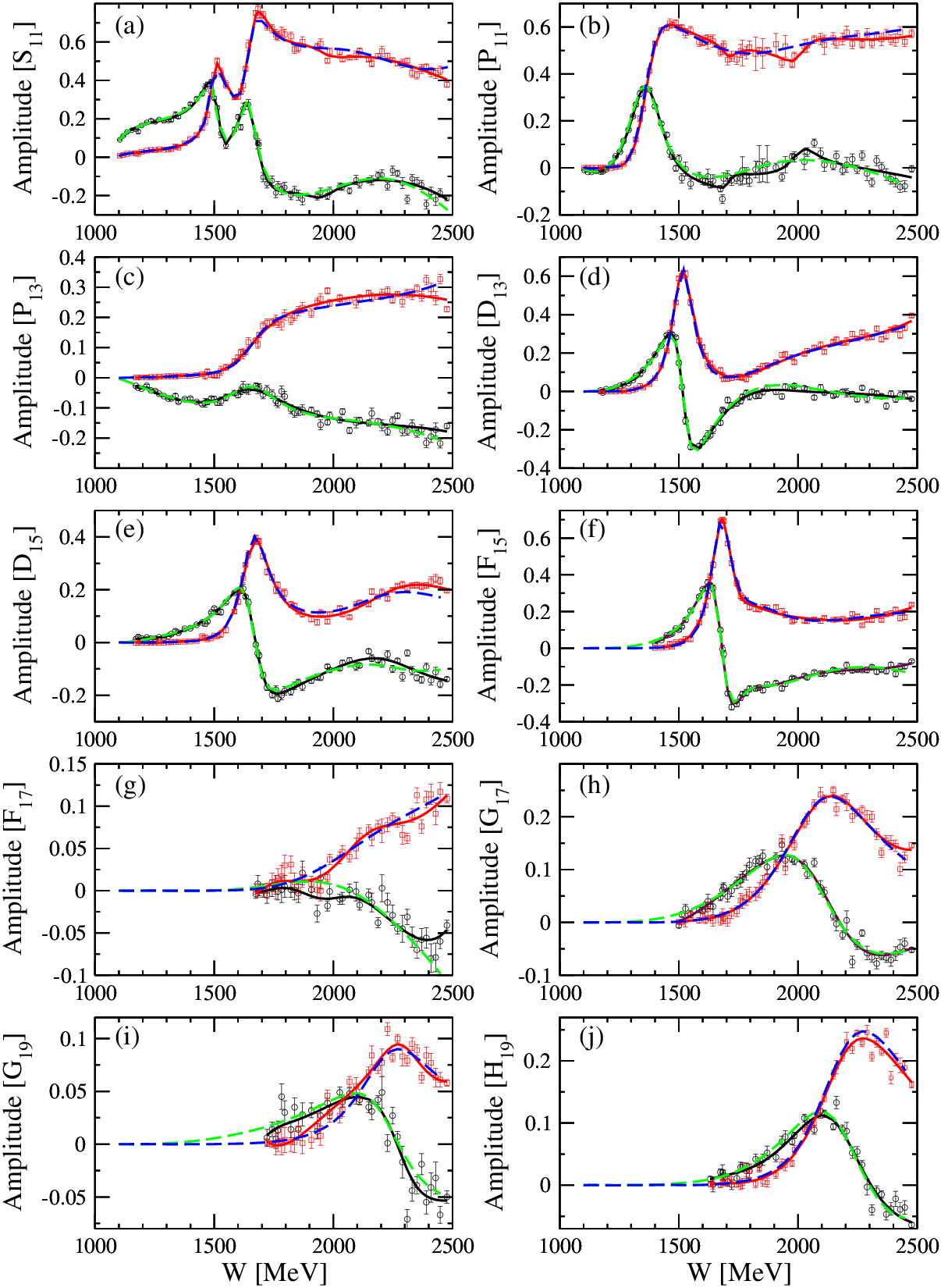}
\caption{(Color online) {\footnotesize Partial-wave amplitudes of the isospin $I=1/2$. Dashed (green and blue) curves give L+P fit to the real (imaginary) parts of the WI08 energy-dependent  
solution. Input data from Ref.~\cite{GWU2}, which are fitted, are not shown as the fitted curve goes strictly through them (see extremely low $D_{dp}$ in Table~\ref{tab:pole1LPpar}  of 
Appendix II).  Solid (black and red) curves give L+P fit to the real (imaginary) parts of amplitudes corresponding to the WI08 SE solution.
(a) $S_{11}$, (b) $P_{11}$, (c) $P_{13}$, (d) $D_{13}$, (e) $D_{15}$, (f) $F_{15}$, (g) $F_{17}$, (h) $G_{17}$, (i) $G_{19}$, and (j) $H_{19}$. All amplitudes are dimensionless.}}
\label{Fig1}
\end{figure*}

\begin{figure*}[!p]
\includegraphics[width=0.88\textwidth]{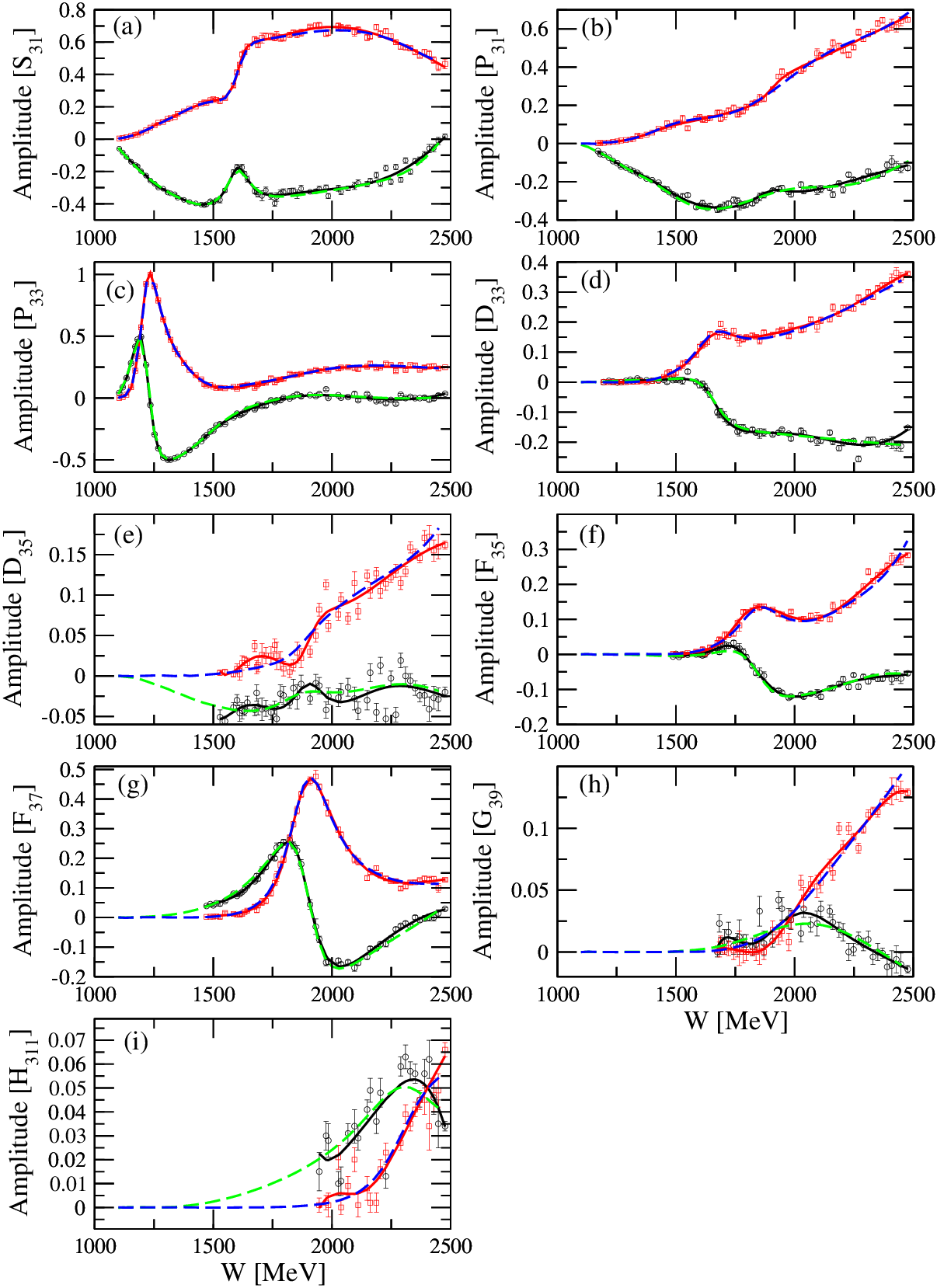}
\caption{(Color online) {\footnotesize Partial-wave amplitudes of the isospin $I=3/2$. Dashed (green and blue) curves give L+P fit to the real (imaginary) parts of the WI08 energy-dependent
solution. Input data from Ref.~\cite{GWU2}, which are fitted, are not shown as the fitted curve goes strictly through them (see extremely low $D_{dp}$ in Table~\ref{tab:pole1LPpar} of 
Appendix II).  Solid (black and red) curves give L+P fit to the real (imaginary) parts of amplitudes corresponding to the WI08 SE solution.
(a) $S_{31}$, (b) $P_{31}$, (c) $P_{33}$, (d) $D_{33}$, (e) $D_{35}$, (f) $F_{35}$, (g) $F_{37}$, (h) $G_{39}$, 
and (i) $H_{3 \; 11}$. All amplitudes are dimensionless.}}
\label{Fig2}
\end{figure*}

We give our main results in Tables~\ref{tab:pole1} and  \ref{tab:pole1a} for I$=1/2$ and in Table~\ref{tab:pole2} for I$=3/2$ partial waves. The agreement of the fit with the input amplitudes  
for I$=1/2$ is given in Fig.~\ref{Fig1} and for I$=3/2$ in Fig.~\ref{Fig2}. Pietarinen expansion parameters for all fitted ED and SE solutions (all branch points $x_P$, $x_Q$, $x_R$,  
number of resonances and discrepancy parameters for both isospins are given in Tables~\ref{tab:pole1LPpar} and \ref{tab:pole2LPpar} of the Appendix.
In Tables \ref{tab:pole1}, \ref{tab:pole1a} and \ref{tab:pole2} we list all resonant states we have found, but attribute an asterisk to those which can be explained by the third branch point becoming 
complex. Alternative solutions with one resonance less are given in Table \ref{tab:pole5}.   
\\ \\ \noindent
First lines (bold-face) give the RPP pole values as taken from Ref.~\cite{PDG}. Second lines, also in bold-face, give the pole values obtained by analytic continuation of T-matrices of WI08 GWU-SAID model 
into the complex energy plane, see Refs.~\cite{GWU,GWU1}. If compared to the published values for SP06 \cite{PDG,GWU} one finds only minor deviations.
The next two lines in these tables show results which were obtained when the L+P method was used to fit the WI08 ED and SE solutions.
\\ \\  \indent From Table~\ref{tab:pole1LPpar} of the Appendix (extremely low $D_{dp}$), it is clear that the L+P formalism is fitting the 
WI08 ED solution almost perfectly for all partial waves and both isospins, thus the input data for WI08 are overlapping with the fitting lines, and are not given in Figs.~\ref{Fig1} and  
\ref{Fig2}. The L+P formalism is therefore 
giving an analytic solution which is consistent with the input WI08 amplitude set.
We generally find that the differences between 
contour-integral and L+P pole parameters are fairly small; on the level of one standard deviation for pole positions and somewhat worse for the residua. 
\\ \\ \noindent
The SE solutions are less dependent on the underlying fit function than are the ED results as discussed above. 
This additional freedom comes at the price
of sacrificing analyticity constraints used to generate the ED fit. The data in single-energy
bins are not sufficient to generate a unique set of partial-wave amplitudes. [For this reason, dispersion-relation constraints
were imposed in the Karlsruhe-Helsinki~\cite{Hoehler84}, Carnegie-Mellon-Berkeley~\cite{cmb}, and the present WI08 analyses.]
Introducing L+P formalism enables us to fit the SE partial-wave results with analytic functions (analyticity is restored) without introducing any explicit model \cite{LP1}. From  
Table~\ref{tab:pole2LPpar} of the Appendix (reasonably low $\chi^2_{dp}$) and from good agreement with the input amplitudes, depicted in Figs.~\ref{Fig1} and \ref{Fig2}, we conclude that L+P  
formalism is fitting SE solution for all partial waves and both isospins reasonably well. 
\\ \\ \indent
When we apply our L+P formalism with real branch points to SE solutions for both isospins, we see that the number of resonances needed to explain the input is increased with respect to the ED  
WI08 result. 
We have found new resonances in S$_{11}$, 
P$_{11}$ (already reported in Ref.~\cite{LP1}), D$_{13}$, F$_{17}$, and S$_{31}$ partial waves.  The only problem occurs in
high $I=3/2$ partial waves G$_{39}$ and H$_{311}$  where the quality of SE solutions is low and no stable solution with resonances is found.  It is interesting to note that the additional poles found in the 
SE amplitude fit 
 N(1895) $1/2^-$, N(1710) $1/2^+$, N(2100)$1/2^{+}$, N(1700)$3/2^-$,  N(1990) $7/2^+$ and $\Delta (1900)1/2^-$
are similar to the ones found in Refs.~\cite{Hoehler84} and \cite{cmb}.

\begin{table*}[ht]
\caption{{\footnotesize \label{tab:pole1} Pole positions in MeV and residues
of I$=1/2$ partial waves;  moduli are given in MeV and phases in degrees. Results for the WI08 ED solution obtained by using formalism \cite{GWU,GWU1} are given in boldface, and  results obtained from the 
L+P approach for WI08 ED and SE solutions are given in normal font. Resonances marked with a
star indicate resonances that can be alternately explained with a $\rho$N complex branch point.
RPP denotes the range of pole parameters given by the authors of Ref. \cite{PDG}.}}
\begin{tabular}{cc|lllll}
\hline \hline
PW & Resonance & Source & Re W$_{p}$ & -2Im W$_{p}$ & $|$residue$|$ & $\theta$\tabularnewline
\hline \hline
\multirow{14}{*}{$S_{11}$} & \multirow{4}{*}{$N(1535)\:1/2^{-}$ } & \textbf{RPP} & \textbf{$\mathbf{1490-1510}$} & \textbf{$\mathbf{90-170}$} & \textbf{$\mathbf{50\pm20}$} & 
\textbf{$\mathbf{(-15\pm15)^{\circ}}$}\tabularnewline
& &\textbf{WI08} & $\mathbf{1499}$ & $\mathbf{98}$ & $\mathbf{17}$ & $\mathbf{-24^\circ}$\tabularnewline
&  & WI08 ED L+P & $1497\pm8\pm1$ & $85\pm14\pm7$ & $13\pm3\pm1$ & $-(41\pm12\pm4)^{\circ}$\tabularnewline
 &  & WI08 SE L+P & $1507\pm1\pm0$ & $88\pm3\pm1$ & $17\pm0.6\pm0.2$ & $-(22\pm2\pm2)^{\circ}$\tabularnewline
\cline{2-7} 
 & \multirow{4}{*}{$N(1650)\:1/2^{-}$} 
 & \textbf{RPP}    & $\mathbf{1640-1655}$ & $\mathbf{100-135}$ & $\mathbf{40-46}$ & \textbf{$\mathbf{(-75\pm 25)^{\circ}}$}\tabularnewline
& & \textbf{WI08} & $\mathbf{1647}$ & $\mathbf{83}$ & $\mathbf{15}$ & $\mathbf{-74^{\circ}}$\tabularnewline
&  & WI08 ED L+P& $1645\pm1\pm4$ & $94\pm9\pm1$ & $20\pm3\pm1$ & $-(77\pm7\pm2)^{\circ}$\tabularnewline
 &  & WI08 SE L+P & $1654\pm2\pm1$ & $112\pm4\pm4$ & $27\pm1\pm2$ & $-(57\pm2\pm2)^{\circ}$\tabularnewline
\cline{2-7} 
 & \multirow{4}{*}{$N(1895)^* \:1/2^{-}$} 
& \textbf{RPP}& \textbf{$\mathbf{1900-2150}$} & \textbf{$\mathbf{90-479}$} & \textbf{$\mathbf{1-60}$} & $\mathbf{(0-164)^{\circ}}$\tabularnewline
&  &\textbf{WI08 } &\textbf{-} &\textbf{-}  &\textbf{-} & \textbf{-}\tabularnewline
&  & WI08 ED L+P & - & - & - & -\tabularnewline
 &  & WI08 SE L+P& $1950\pm16\pm6$ & $170\pm37\pm23$ & $6\pm1\pm1$ & $(97\pm10\pm5)^{\circ}$\tabularnewline
\hline \hline
\multirow{14}{*}{$P_{11}$} & \multirow{4}{*}{$N(1440)\:1/2^{+}$} & 
\textbf{RPP} &  $\mathbf{1350-1365}$ & $\mathbf{160-190}$ & $\mathbf{40-52}$ & $\mathbf{(-100\pm 35)^{\circ}}$\tabularnewline
& & \textbf{WI08} & $\mathbf{1358}$ & $\mathbf{160}$ & $\mathbf{37}$ & $\mathbf{-98^{\circ}}$\tabularnewline
 &  & WI08 ED L+P & $1358\pm2\pm1$ & $180\pm6\pm1$ & $45\pm1\pm1$ & $-(91\pm1\pm1)^{\circ}$\tabularnewline
 &  & WI08 SE L+P & $1364\pm0.7\pm0.3$ & $182\pm1\pm0.5$ & $45\pm0.4\pm0.3$ & $-(86\pm0.5\pm0.3)^{\circ}$\tabularnewline
\cline{2-7} 
 & \multirow{4}{*}{$N(1710)^*\:1/2^{+}$} &
\textbf{RPP} & $\mathbf{1670-1720}$ & $\mathbf{80-230}$ & $\mathbf{6-15}$ & ${\color{black}{\color{black}\mathbf{(90-200)^{\circ}}}}$\tabularnewline
&&\textbf{WI08} & \textbf{-} & \textbf{-} & \textbf{-} & \textbf{-}\tabularnewline
 &  & WI08 ED L+P & - & - & - & -\tabularnewline
 &  & WI08 SE L+P & $1711\pm10\pm0.6$ & $84\pm20\pm2$ & $2\pm0.7\pm0.1$ & $(171\pm14\pm0.4)^{\circ}$\tabularnewline
\cline{2-7} 
 & \multirow{4}{*}{$N(2100)^*\:1/2^{+}$} 
& \textbf{RPP} & $\mathbf{2120\pm40}$ & $\mathbf{180-420}$ & $\mathbf{14\pm7}$ & $\mathbf{(35\pm25)^{\circ}}$\tabularnewline
& & \textbf{WI08} & \textbf{-} & \textbf{-} & \textbf{-} & \textbf{-}\tabularnewline
 &  & WI08 ED L+P & - & - & - & -\tabularnewline
 &  & WI08 SE L+P & $2004\:\pm10\pm1.3$ & $140\pm20\pm1.2$ & $7\pm0\pm9$ & $-(126\pm22\pm1)^{\circ}$\tabularnewline
\hline \hline
\multirow{4}{*}{$P_{13}$} 
& & \textbf{RPP}   & $\mathbf{1660-1690}$ & $\mathbf{150-400}$ & $\mathbf{15\pm8}$ & $\mathbf{(-130\pm30)^{\circ}}$\tabularnewline
 &\multirow{2}{*}{$N(1720)\:3/2^{+}$}   & \textbf{WI08} & $\mathbf{1661}$ & $\mathbf{304}$ & $\mathbf{21}$ & $\mathbf{-89^{\circ}}$\tabularnewline
 &  & WI08 ED  L+P& $1659\pm10\pm1$ & $303\pm18\pm1$ & $20\pm2\pm1$ & $-(91\pm6\pm1)^{\circ}$\tabularnewline
 &  & WI08 SE L+P & $1668\pm15\pm9$ & $303\pm18\pm40$ & $16\pm1\pm6$ & $-(82\pm4\pm8)^{\circ}$\tabularnewline
\hline \hline
\end{tabular}
\end{table*}
\begin{table*}[ht]
\caption{{\footnotesize \label{tab:pole1a} Pole positions in MeV and residues
of I$=1/2$ partial waves;  moduli are given in MeV and phases in degrees. The results for the WI08 ED solution obtained by using formalism \cite{GWU,GWU1} are given in boldface, and  results obtained from 
the L+P approach for WI08 ED and SE solutions are given in normal font. Resonances marked with a
star indicate resonances that can be explained by $\rho$N complex branch point.
RPP denotes the range of pole parameters given by the authors of Ref. \cite{PDG}.}}
\begin{tabular}{cc|lllll}
\hline \hline
PW & Resonance & Source & Re W$_{p}$ & -2Im W$_{p}$ & $|$residue$|$ & $\theta$\tabularnewline
\hline \hline
\multirow{9}{*}{$D_{13}$} 
&  &\textbf{RPP}  & $\mathbf{1505-1515}$ & $\mathbf{105-120}$ & $\mathbf{35\pm3}$ & $\mathbf{(-10\pm5)^{\circ}}$\tabularnewline
& \multirow{3}{*}{$N(1520)\:3/2^{-}$} & \textbf{WI08} & \textbf{$\mathbf{1515}$} & \textbf{$\mathbf{110}$} & $\mathbf{37}$ & $\mathbf{-4^{\circ}}$\tabularnewline
 &  & WI08 ED L+P & $1515\pm1\pm1$ & $109\pm3\pm1$ & $37\pm1\pm1$ & $-(7\pm1\pm1)^{\circ}$\tabularnewline
 &  & WI08 SE L+P & $1512\pm1\pm1$ & $113\pm2\pm4$ & $37\pm1\pm2$ & $-(8\pm1\pm1)^{\circ}$\tabularnewline
\cline{2-7} 
 & \multirow{4}{*}{$N(1700)^*\:3/2^{-}$}
& \textbf{RPP}   & $\mathbf{1650-1750}$ & $\mathbf{100-350}$ & $\mathbf{5-50}$ & $\mathbf{(-120\: to\:20)^{\circ}}$\tabularnewline
 & & \textbf{WI08} & \textbf{-} & \textbf{-} & \textbf{-} & \textbf{-}\tabularnewline
 &  & WI08 ED L+P & - & - & - & -\tabularnewline
 &  & WI08 SE L+P & $1752\pm8\pm9$ & $572\pm16\pm19$ & $49\pm1\pm4$ & $-(121\pm2\pm7)^{\circ}$\tabularnewline
\hline \hline
\multirow{5}{*}{$D_{15}$} 
&  &\textbf{RPP}  & $\mathbf{1655-1665}$ & $\mathbf{125-150}$ & $\mathbf{25\pm5}$ & $\mathbf{(-25\pm6){}^{\circ}}$\tabularnewline
& \multirow{3}{*}{$N(1675)\:5/2^{-}$} & \textbf{WI08} & $\mathbf{1656}$ & $\mathbf{140}$ & $\mathbf{27}$ & $\mathbf{-22^{\circ}}$\tabularnewline
 &  & WI08 ED L+P & $1657\pm2\pm1$ & $139\pm4\pm1$ & $27\pm1\pm1$ & $-(20\pm1\pm1)^{\circ}$\tabularnewline
 &  & WI08 SE L+P & $1661\pm1\pm0$ & $147\pm2\pm0.4$ & $28\pm0.4\pm0.2$ & $-(18\pm1\pm0)^{\circ}$\tabularnewline
\hline \hline
\multirow{10}{*}{$F_{15}$} 
& &\textbf{RPP} & $\mathbf{1665-1680}$ & $\mathbf{110-135}$ & $\mathbf{40\pm5}$ & $\mathbf{(-10\pm10){}^{\circ}}$\tabularnewline
& \multirow{3}{*}{$N(1680)\:5/2^{+}$} & \textbf{WI08} & \textbf{$\mathbf{1674}$} & \textbf{$\mathbf{113}$} & $\mathbf{40}$ & $\mathbf{-3^{\circ}}$\tabularnewline
 &  & WI08 ED L+P & $1674\pm2\pm1$ & $114\pm5\pm2$ & $42\pm3\pm2$ & $-(9\pm3\pm2)^{\circ}$\tabularnewline
 &  & WI08 SE L+P & $1678\pm1\pm3$ & $113\pm1\pm2$ & $43\pm1\pm3$ & $(-3\pm1\pm4)^{\circ}$\tabularnewline
\cline{2-7} 
 & \multirow{5}{*}{$N(1860)^*\:5/2^{+}$} 
& \textbf{RPP}& $\begin{array}{c}
\mathbf{2030\pm110}\\
\mathbf{or\;1779}
\end{array}$ & $\begin{array}{c}
\mathbf{480\pm100}\\
\mathbf{or\:248}
\end{array}$ & $\mathbf{50\pm 20}$ &   
$\begin{array}{c}
\mathbf{(-80\pm 40)^{\circ }}\\

\end{array}$
\tabularnewline
& & \textbf{WI08} & \textbf{$\mathbf{1779}$} & \textbf{$\mathbf{275}$} & ${\mathbf{53}}$ & $\mathbf{-63^{\circ}}$\tabularnewline
 &  & WI08 ED L+P & $1774\pm20\pm10$ & $242\pm38\pm10$ & $28\pm4\pm4$ & $-(70\pm9\pm18)^{\circ}$\tabularnewline
 &  & WI08 SE L+P & $1794\pm1\pm28$ & $212\pm1\pm26$ & $18\pm1\pm3$ & $-(63\pm4\pm27)^{\circ}$\tabularnewline
\hline \hline
\multirow{4}{*}{$F_{17}$} 
& &\textbf{RPP}& $\begin{array}{c}
\mathbf{2030\pm69}\\
\mathbf{1900\pm 30}
\end{array}$ & $\begin{array}{c}
\mathbf{240\pm60}\\
\mathbf{260\pm 60}
\end{array}$ &  $\begin{array}{c}
\mathbf{2\pm 1}\\
\mathbf{9\pm 3}
\end{array}$ & $\begin{array}{c}
\mathbf{(125\pm65)^{\circ}}\\
\mathbf{(-60\pm 30)^{\circ }}
\end{array}$ \tabularnewline
& \multirow{2}{*}{$N(1990)\:7/2^{+}$} & \textbf{WI08} & \textbf{-} & \textbf{-} & \textbf{-} & \textbf{-}\tabularnewline
 &  & WI08 ED L+P & - & - & - & -\tabularnewline
 &  & WI08 SE L+P & $2157\pm32\pm30$ & $261\pm64\pm40$ & $4\pm2\pm1$ & $-(33\pm21\pm15)^{\circ}$\tabularnewline
\hline \hline
\multirow{4}{*}{$G_{17}$} 

& \multirow{4}{*}{$N(2190)\:7/2^{-}$} 
 & \textbf{RPP} & $\mathbf{2050-2100}$ & $\mathbf{400-520}$ & $\mathbf{30-72}$ & $\mathbf{(-30\: to\:30)^{\circ}}$\tabularnewline
& &\textbf{WI08} &$ \mathbf{2066}$ & $\mathbf{534}$ & $\mathbf{74}$ & $\mathbf{-32^{\circ}}$\tabularnewline
 &  & WI08 ED L+P & $2060\pm4\pm 7$ & $521\pm10\pm 6$ & $69\pm2\pm2$ & $-(36\pm2\pm4)^{\circ}$\tabularnewline
 &  & WI08 SE L+P & $2132\pm5\pm19$ & $550\pm11\pm14$ & $82\pm2\pm5$ & $-(11\pm 2\pm2
)^{\circ}$\tabularnewline
\hline \hline
\multirow{4}{*}{$G_{19}$}
 & &\textbf{RPP}& \textbf{$\mathbf{2150-2250}$} & \textbf{$\mathbf{350-550}$} & \textbf{$\mathbf{20-30}$} & $\mathbf{(-50\pm30)^{\circ}}$\tabularnewline
 & \multirow{2}{*}{$N(2250)\:9/2^{-}$} & \textbf{WI08} &$ \mathbf{2221}$ & $\mathbf{416}$ & $\mathbf{19}$ & $\mathbf{-28^{\circ}}$\tabularnewline
 &  & WI08 ED L+P & $2224\pm4\pm1$ & $417\pm9\pm1$ & $19\pm1\pm1$ & $-(26\pm1\pm1)^{\circ}$\tabularnewline
 &  & WI08 SE L+P & $2283\pm10\pm0$ & $304\pm20\pm11$ & $12\pm1\pm2$ & $(14\pm5\pm2)^{\circ}$\tabularnewline
\hline \hline
\multirow{4}{*}{$H_{19}$} & \multirow{4}{*}{$N(2220)\:9/2^{+}$}
 &\textbf{RPP} & $\mathbf{2130-2200}$ & $\mathbf{400-560}$ & $\mathbf{33-60}$ & $\mathbf{(-45\pm25)^{\circ}}$\tabularnewline
 & & \textbf{WI08} &$ \mathbf{2169}$ & $\mathbf{488}$ & $\mathbf{63}$ & $\mathbf{-56^{\circ}}$\tabularnewline
 &  & WI08 ED L+P & $2177\pm3\pm1$ & $464\pm8\pm1$ & $56\pm1\pm1$ & $-(50\pm1\pm2)^{\circ}$\tabularnewline
 &  & WI08 SE L+P & $2173\pm5\pm2$ & $445\pm11\pm10$ & $49\pm2\pm3$ & $-(53\pm2\pm2)^{\circ}$\tabularnewline
\hline \hline
\end{tabular}
\end{table*}


\begin{table*}[ht]
\caption{{\footnotesize \label{tab:pole2} Pole positions in MeV and residues
of I$=3/2$ partial waves;  moduli are given in MeV and phases in degrees. The results for the WI08 ED solution obtained by using formalism \cite{GWU,GWU1} are given in boldface, and  results obtained from 
the L+P approach are for WI08 ED and SE solutions  given in normal font. Resonances marked with a
star indicate resonances that can be explained by $\rho$N complex branch point.
RPP denotes the range of pole parameters given by the authors of Ref. \cite{PDG}.
}}
\begin{tabular}{cc|lllll}
\hline \hline
PW & \multicolumn{1}{c}{Resonance} & Source & Re W$_{p}$ & -2Im W$_{p}$ & $|$residue$|$ & $\theta$\tabularnewline
\hline \hline
\multirow{10}{*}{$S_{31}$} 
& & \textbf{RPP}& $\mathbf{1590-1610}$ & $\mathbf{120-140}$ & $\mathbf{13-20}$ & $\mathbf{(-110\pm20)^{\circ}}$\tabularnewline
& \multirow{3}{*}{$\Delta(1620)\:1/2^{-}$} & \textbf{WI08} & \textbf{$\mathbf{1594}$} & \textbf{$\mathbf{136}$} & $\mathbf{18}$ & $\mathbf{-108^{\circ}}$\tabularnewline
 &  & WI08 ED L+P & $1595\pm9\pm4$ & $138\pm9\pm6$ & $18\pm3\pm1$ & $-(107\pm10\pm9)^{\circ}$\tabularnewline
 &  & WI08 SE L+P & $1599\pm2\pm4$ & $114\pm4\pm3$ & $15\pm1\pm1$ & $-(102\pm2\pm9)^{\circ}$\tabularnewline
\cline{2-7}
 & \multirow{4}{*}{$\Delta(1900)^*\:1/2^{-}$} 
& \textbf{RPP} & ${\begin{array}{c}
\mathbf{1820-1910}\\
\mathbf{or}\:\mathbf{1780}
\end{array}}$ & $\mathbf{130-345}$ & $\mathbf{10\pm3}$ & $\begin{array}{c}
\mathbf{(-125\pm20)^{\circ}}\\
\mathbf{or\:(20\pm40)^{\circ}}
\end{array}$\tabularnewline
& & \textbf{WI08 } & \textbf{-} & \textbf{-} & \textbf{-} & \textbf{-}\tabularnewline
 &  & WI08 ED L+P&-  &-  &-  &- \tabularnewline
 &  & WI08 SE L+P & $1878\pm20\pm22$ & $160\pm40\pm40$ & $6\pm1\pm4$ & $-(45\pm17\pm22)^{\circ}$\tabularnewline
\hline \hline
\multirow{5}{*}{$P_{31}$}
& &\textbf{RPP}& $\mathbf{1830-1880}$ & $\mathbf{200-500}$ & $\mathbf{16-45}$ & $\mathbf{-}$\tabularnewline
 & \multirow{3}{*}{$\Delta(1910)\:1/2^{+}$} & \textbf{WI08} & $\mathbf{1764}$ & $\mathbf{478}$ & $\mathbf{46}$ & $\mathbf{171^{\circ}}$\tabularnewline
 &  & WI08 ED L+P& $1744\pm22\pm1$ & $361\pm43\pm1$ & $24\pm3\pm1$ & $(153\pm8\pm1)^{\circ}$\tabularnewline
 &  & WI08 SE L+P & $1773\pm16\pm2$ & $326\pm24\pm2$ & $18\pm2\pm1$ & $(176\pm7\pm1)^{\circ}$\tabularnewline
\hline \hline
\multirow{10}{*}{$P_{33}$}
 && \textbf{RPP}  & $\mathbf{1209-1211}$ & $\mathbf{98-102}$ & $\mathbf{50\pm3}$ & $(-\mathbf{46\pm2)^{\circ}}$\tabularnewline
 & \multirow{3}{*}{$\Delta(1232)\:3/2^{+}$} & \textbf{WI08} & $\mathbf{1211}$ & $\mathbf{100}$ & $\mathbf{53}$ & $\mathbf{-47^{\circ}}$\tabularnewline
 &  & WI08 ED L+P& $1211\pm1\pm1$ & $98\pm2\pm1$ & $49\pm2\pm2$ & $-(46\pm2\pm2)^{\circ}$\tabularnewline
 &  & WI08 SE L+P& $1211\pm0\pm0$ & $100\pm1\pm1$ & $52\pm1\pm0$ & $-(46\pm0\pm0)^{\circ}$\tabularnewline
\cline{2-7} 
 & \multirow{3}{*}{$\Delta(1600)\:3/2^{+}$}
& \textbf{RPP} & $\mathbf{1460-1560}$ & $\mathbf{200-350}$ & $\mathbf{5-44}$ & $\mathbf{-}$\tabularnewline
 & & \textbf{WI08} & $\mathbf{1457}$ & $\mathbf{421}$ & \textbf{$\mathbf{52}$} & $\mathbf{153^{\circ}}$\tabularnewline
 &  & WI08 ED L+P & $1505\pm19\pm20$ & $449\pm31\pm54$ & $54\pm4\pm4$ & $(164\pm3\pm8)^{\circ}$\tabularnewline
 &  & WI08 SE L+P & $1493\pm7\pm9$ & $245\pm8\pm46$ & $45\pm10\pm27$ & $-(135\pm10\pm18)^{\circ}$\tabularnewline
\hline \hline
\multirow{5}{*}{$D_{33}$} 
& & \textbf{RPP} & $\mathbf{1620-1680}$ & $\mathbf{160-300}$ & $\mathbf{10-50}$ & $\mathbf{(-45\: to\:12)^{\circ}}$\tabularnewline
& \multirow{3}{*}{$\Delta(1700)\:3/2^{-}$} & \textbf{WI08} & $\mathbf{1649}$ & $\mathbf{258}$ & $\mathbf{18}$ & $\mathbf{-22^{\circ}}$\tabularnewline
 &  & WI08 ED L+P & $1652\pm7\pm3$ & $248\pm14\pm14$ & $17\pm1\pm3$ & $-(16\pm4\pm2)^{\circ}$\tabularnewline
 &  & WI08 SE L+P& $1646\pm4\pm7$ & $203\pm8\pm9$ & $12\pm1\pm2$ & $-(13\pm5\pm8)^{\circ}$\tabularnewline
\hline \hline
\multirow{5}{*}{$D_{35}$}
& &\textbf{RPP} & $\mathbf{1840-1960}$ & $\mathbf{175-360}$ & $\mathbf{7-30}$ & $\mathbf{(-20\pm40)^{\circ}}$\tabularnewline
 & \multirow{3}{*}{$\Delta(1930)\:5/2^{-}$} & \textbf{WI08} & $\mathbf{1970}$ & $\mathbf{292}$ & $\mathbf{5}$ & $\mathbf{-19^{\circ}}$\tabularnewline
 &  & WI08 ED L+P& $1969\pm15\pm8$ & $248\pm32\pm4$ & $4\pm1\pm1$ & $-(13\pm8\pm5)^{\circ}$\tabularnewline
 &  & WI08 SE L+P& $1845\pm7\pm24$ & $174\pm15\pm25$ & $4\pm1\pm2$ & $-(163\pm5\pm26)^{\circ}$\tabularnewline
\hline \hline
\multirow{5}{*}{$F_{35}$} 
 & &\textbf{RPP} & $\mathbf{1805-1835}$ & $\mathbf{265-300}$ & $\mathbf{25\pm10}$ & $\mathbf{(-50\pm20)^{\circ}}$ \tabularnewline
& \multirow{3}{*}{$\Delta(1905)\:5/2^{+}$} & \textbf{WI08} & $\mathbf{1817}$ & $\mathbf{252}$ & $\mathbf{15}$ & $\mathbf{-31^{\circ}}$\tabularnewline
 &  & WI08 ED L+P& $1814\pm4\pm1$ & $273\pm7\pm2$ & $18\pm1\pm1$ & $-(34\pm2\pm1)^{\circ}$\tabularnewline
 &  & WI08 SE L+P& $1831\pm4\pm3$ & $329\pm6\pm11$ & $29\pm1\pm3$ & $-(8\pm2\pm4)^{\circ}$\tabularnewline
\hline \hline
\multirow{5}{*}{$F_{37}$} 
& &\textbf{RPP}  & $\mathbf{1870-1890}$ & $\mathbf{220-260}$ & $\mathbf{47-61}$ & $\mathbf{(-33\pm12)^{\circ}}$\tabularnewline
& \multirow{3}{*}{$\Delta(1950)\:7/2^{+}$} & \textbf{WI08} & \textbf{$\mathbf{1883}$} & \textbf{$\mathbf{231}$} & $\mathbf{55}$ & $\mathbf{-25^{\circ}}$\tabularnewline
 &  & WI08 ED L+P& $1878\pm2\pm2$ & $227\pm5\pm1$ & $53\pm1\pm1$ & $-(22\pm1\pm1)^{\circ}$\tabularnewline
 &  & WI08 SE L+P & $1888\pm1\pm2$ & $234\pm3\pm3$ & $57\pm1\pm1$ & $-(31\pm1\pm2)^{\circ}$\tabularnewline
\hline \hline
\multirow{4}{*}{$G_{39}$} &
 \multirow{4}{*}{$\Delta(2400)\:9/2^{-}$} 
&\textbf{RPP}  & $\mathbf{1983}$ & $\mathbf{878}$ & $\mathbf{24}$ & $\mathbf{-139^{\circ}}$\tabularnewline
& & \textbf{WI08} & \textbf{$\mathbf{1966}$} & \textbf{$\mathbf{855}$} & $\mathbf{22}$ & $\mathbf{-135^{\circ}}$\tabularnewline
 &  & WI08 ED L+P& $1955\pm7\pm17$ & $911\pm13\pm11$ & $27\pm1\pm2$ & $-(149\pm1\pm8)^{\circ}$\tabularnewline
 &  & WI08 SE L+P & - & - & - & -\tabularnewline
\hline \hline
\multirow{5}{*}{$H_{311}$} 
 & &\textbf{RPP}  & $\mathbf{2260-2400}$ & $\mathbf{350-750}$ & $\mathbf{12-39}$ & $\mathbf{-(30\pm40)^{\circ}}$\tabularnewline
& \multirow{3}{*}{$\Delta(2420)\:11/2^{+}$} & \textbf{WI08} & \textbf{$\mathbf{2316}$} & \textbf{$\mathbf{448}$} & $\mathbf{7}$ & $\mathbf{-73^{\circ}}$\tabularnewline
 &  & WI08 ED L+P& $2320\pm9\pm4$ & $442\pm16\pm7$ & $7\pm0.3\pm0.2$ & $-(71\pm3\pm3)^{\circ}$\tabularnewline
 &  & WI08 SE L+P & - & - & - & -\tabularnewline
\hline \hline
\end{tabular}
\end{table*}


\newpage

\clearpage
\subsection{Complex branch points}
As mentioned in Refs.\cite{LP1,LP1}, the use of a complex branch points, applied to ED and SE fits, can produce a result
with one less pole. This test was also applied here, assuming a $\rho N$ branch point, and was capable of compensating for
the additional pole appearing in the $S_{11}$, $P_{11}$, $D_{13}$, $F_{15}$, and $S_{31}$ partial waves.
This issue cannot be resolved from elastic
pion-nucleon scattering data alone, as it involves the coupling to final states beyond $\pi N$, such
as $\pi \Delta$, $\rho N$, and $\sigma N$.
\\ \\ \indent
In Table \ref{tab:pole5} we give the parameters for some typical situations when fits with complex branch point achieve a similar quality as fits with  real branch points (measured by the size of a 
discrepancy variable $D_{dp}$, see Eq. \ref{discrepancy} ). 
Here we repeat that the complex branch point is a mathematical implementation of the situation when the three-body final state contains a two-body  resonant sub-channel accompanied by the third ``observer" 
particle. So, we effectively only replace a resonance contribution from an isobar intermediate state by a resonant contribution in a three body sub-channel (final state interaction).
It is important to accept the fact that both mechanisms (real and complex branch points) are, unfortunately, indistinguishable in a single-channel model. 
As was the case in  the J\"{u}lich model for $P_{11}$(1710), other channels (K$\Lambda$ channel in the J\"{u}lich model) are essential to distinguish between the two alternatives. 
We have further attempted to use complex branch points and additional resonances in fit the data.  
However, without knowing the branching fraction of two-body to three-body channel, the complex branch point takes over the whole flux, and eliminates the additional resonances altogether.

\begin{table*}[ht]
{\footnotesize \caption{{\footnotesize \label{tab:pole5} Pole positions in MeV and residues
of I$=1/2$ and I$=1/2$ partial waves;  moduli are given in MeV and phases in degrees.  $N_r$ is  number of resonance poles.
The results from L+P expansion are given for WI08 ED and SE  solutions using a $\rho N$ complex
branch point.}}
}
{\footnotesize \par}

\begin{tabular}{ccc|ccccccccc}
\hline 
PW & Source  & $N_{r}$ & \textcolor{black}{Resonance}  & \textcolor{black}{$\;\mbox{Re}\, W_{p}\;$ }  & \textcolor{black}{$\;-2\mbox{Im}\, W_{p}\;$ }  & \textcolor{black}{$|\mbox{residue}|$ }  & 
\textcolor{black}{$\theta$ }  & $x_{P}$ & $x_{Q}$ & $x_{R}$ & $\chi^{2}$\tabularnewline
\hline 
\hline 
$S_{11}$ & WI08 SE & 2 & $N(1535)\:1/2^{-}$ & $1511$ & $83$ & $16$ & $-12^{\circ}$ & $743$ & $1077^{\pi N}$ & $(1700-70i)^{\rho N}$ & $2.77$\tabularnewline
 &  &  & $N(1650)\:1/2^-$ & $1665$ & $109$ & $27$ & $-34^{\circ}$ &  &  &  & \tabularnewline
\hline 
\hline 
$P_{11}$  & WI08 SE  & 1 & $N(1440)\:1/2^{+}$  & $1364$  & $181$  & $44$  & $-85^{\circ}$ & $-276$ & $1077^{\pi N}$ & $(1700-70i)^{\rho N}$ & $1.42$\tabularnewline
\hline \hline
$D_{13}$ & WI08 SE & 1 & \textcolor{black}{$N(1520)\;3/2^{-}$}  & $1512$  & $108$  & $34$  & $-7{}^{\circ}$ & $900$ & $1077^{\pi N}$ & $(1700-70i)^{\rho N}$ & $1.90$\tabularnewline
\hline \hline
$F_{15}$ & WI08 ED  & 1 & \textcolor{black}{$N(1680)\;5/2^{+}$}  & $1672$  & $121$  & $46$  & $-15{}^{\circ}$ & $178$ & $1077^{\pi N}$ & $(1700-70i)^{\rho N}$ & $0.0014$\tabularnewline

 & WI08 SE  & 1 &  & $1679$  & $128$  & $52$  & $-33{}^{\circ}$ & $900$ & $1077^{\pi N}$ & $(1700-70i)^{\rho N}$ & $2.79$\tabularnewline
\hline \hline
$S_{31}$ & WI08 SE & 1 & $\Delta(1620)\:1/2^{-}$ & $1601$ & $113$ & $16$ & $-100^{\circ}$ & $-526$ & $1077^{\pi N}$ & $(1700-70i)^{\rho N}$ & $1.72$\tabularnewline
\hline \hline 
\end{tabular}
\end{table*}

\section{Conclusions}

We have investigated the pole structure of the most recent GWU-SAID partial-wave analysis of pion-nucleon elastic scattering
and eta-nucleon production data (WI08). For the ED fit results, both the use of contour integrals and the L+P expansion method produce
pole positions and residues which are generally consistent within the estimated L+P uncertainties. Given this success, the L+P
expansion has been applied to the descrete SE fits associated with the WI08 solution. This provides an analytic form which again
produces poles and residues in a case where the contour method is not applicable. Having pole parameters with error estimates, 
for both the ED and SE fits, facilitates a comparison of our results with those now becoming available from other groups.
\\ \\ \indent
Systematic deviations from the ED fits, seen in the discrete SE 
amplitudes, require additional poles to achieve a good description. One can generally associate these added poles with
structures seen in the earlier CMB fit~\cite{cmb} and a more recent fit~\cite{Bonn} by the Bonn-Gatchina group. 
As noted in Ref.~\cite{LP1}, some of these structures can also be accounted for, in the L+P method, 
through the introduction of complex branch points. This ambiguity remains if only the elastic scattering data is analyzed.

\clearpage

\section*{appendix}

\begin{table*}[ht]
\caption{\footnotesize \label{tab:pole1LPpar} Parameters from L+P expansion are given for WI08 ED solutions. 
$N_r$ is number of resonance poles, $x_P, x_Q, x_R$ are branch points in $MeV$.}
\begin{tabular}{c|crcll|ccrcll}
\hline \hline
\multicolumn{12}{c}{Source}\tabularnewline
\multicolumn{12}{c}{WI08 ED L+P}\tabularnewline
\hline 
\hline 
PW & $N_{r}$ & $x_{P}$ & $x_{Q}$ & $x_{R}$ & $10^{2}D_{dp}$ & PW & $N_{r}$ & $x_{P}$ & $x_{Q}$ & $x_{R}$ & $10^{2}D_{dp}$\tabularnewline
\hline \hline
$S_{11}$ & $2$ & $900$ & $1077^{\pi N}$ & $1215^{\pi\pi N}$ & $0.91$ & $S_{31}$ & $1$ & $884$ & $1077^{\pi N}$ & $1215^{\pi\pi N}$ & $0.01$\tabularnewline
 & $2$ & $900$ & $1077^{\pi N}$ & $1486^{\eta N}$ & $0.42$ &  & $1$ & $841$ & $1077^{\pi N}$ & $1370^{Real(\pi\Delta)}$ & $0.02$\tabularnewline
 & $2$ & $900$ & $1077^{\pi N}$ & $1480^{free}$ & $0.40$ &  & $1$ & $899$ & $1077^{\pi N}$ & $1708^{Real(\rho N)}$ & $0.05$\tabularnewline
 &  &  &  &  &  &  & $1$ & $834$ & $1077^{\pi N}$ & $1633{}^{free}$ & $0.01$\tabularnewline
\hline \hline
$P_{11}$ & $1$ & $292$ & $1077^{\pi N}$ & $1215^{\pi\pi N}$ & $0.59$ & $P_{31}$ & $1$ & $684$ & $1077^{\pi N}$ & $1215^{\pi\pi N}$ & $0.88$\tabularnewline
 & $1$ & $206$ & $1077^{\pi N}$ & $1370^{Real(\pi\Delta)}$ & $0.47$ &  & $1$ & $623$ & $1077^{\pi N}$ & $1250^{free}$ & $0.87$\tabularnewline
 & $1$ & $401$ & $1077^{\pi N}$ & $1381^{free}$ & $0.40$ &  &  &  &  &  & \tabularnewline
\hline \hline
$P_{13}$ & $1$ & $183$ & $1077^{\pi N}$ & $1215^{\pi\pi N}$ & $0.05$ & $P_{33}$ & $2$ & $-273$ & $1077^{\pi N}$ & $1215^{\pi\pi N}$ & $0.11$\tabularnewline
 & $1$ & $172$ & $1077^{\pi N}$ & $1370^{Real(\pi\Delta)}$ & $0.05$ &  & $2$ & $747$ & $1077^{\pi N}$ & $1370^{Real(\pi\Delta)}$ & $0.07$\tabularnewline
 & $1$ & $-298$ & $1077^{\pi N}$ & $1318^{free}$ & $0.04$ &  & $2$ & $8$ & $1077^{\pi N}$ & $1280{}^{free}$ & $0.04$\tabularnewline
\hline 
\hline 
$D_{13}$ & $1$ & $-809$ & $1077^{\pi N}$ & $1215^{\pi\pi N}$ & $0.43$ & $D_{33}$ & $1$ & $779$ & $1077^{\pi N}$ & $1215^{\pi\pi N}$ & $0.07$\tabularnewline
 & $1$ & $-1831$ & $1077^{\pi N}$ & $1370^{Real(\pi\Delta)}$ & $0.22$ &  & $1$ & $-149$ & $1077^{\pi N}$ & $1370^{Real(\pi\Delta)}$ & $0.06$\tabularnewline
 & $1$ & $44$ & $1077^{\pi N}$ & $1708^{Real(\rho N)}$ & $0.67$ &  & $1$ & $-371$ & $1077^{\pi N}$ & $1284{}^{free}$ & $0.05$\tabularnewline
 & $1$ & $76$ & $1077^{\pi N}$ & $1277{}^{free}$ & $0.15$ &  &  &  &  &  & \tabularnewline
\hline 
\hline 
$D_{15}$ & $1$ & $314$ & $1077^{\pi N}$ & $1215^{\pi\pi N}$ & $0.29$ & $D_{35}$ & $1$ & $472$ & $1077^{\pi N}$ & $1688^{K\Sigma}$ & $0.95$\tabularnewline
 & $1$ & $-1948$ & $1077^{\pi N}$ & $1370^{Real(\pi\Delta)}$ & $0.41$ &  & $1$ & $234$ & $1077^{\pi N}$ & $1345^{free}$ & $0.45$\tabularnewline
 & $1$ & $699$ & $1077^{\pi N}$ & $1340{}^{free}$ & $0.24$ &  &  &  &  &  & \tabularnewline
\hline \hline
$F_{15}$ & $2$ & $232$ & $1077^{\pi N}$ & $1215^{\pi\pi N}$ & $0.04$ & $F_{35}$ & 1 & $181$ & $1077^{\pi N}$ & $1215^{\pi\pi N}$ & $0.55$\tabularnewline
 & $2$ & $333$ & $1077^{\pi N}$ & $1370^{Real(\pi\Delta)}$ & $0.05$ &  & $1$ & $-867$ & $1077^{\pi N}$ & $1708^{Real(\rho N)}$ & $0.58$\tabularnewline
 & $2$ & $601$ & $1077^{\pi N}$ & $1708^{Real(\rho N)}$ & $0.04$ &  & $1$ & $-656$ & $1077^{\pi N}$ & $1225{}^{free}$ & $0.54$\tabularnewline
 & $2$ & $-483$ & $1077^{\pi N}$ & $1691{}^{free}$ & $0.04$ &  &  &  &  &  & \tabularnewline
\hline \hline
$F_{17}$ &  &  &  &  &  & $F_{37}$ & 1 & $-202$ & $1077^{\pi N}$ & $1370^{Real(\pi\Delta)}$ & $0.14$\tabularnewline
 &  & \multicolumn{3}{c}{No results} &  &  & 1 & $-918$ & $1077^{\pi N}$ & $1708^{Real(\rho N)}$ & $0.16$\tabularnewline
 &  &  &  &  &  &  & 1 & $-38$ & $1077^{\pi N}$ & $1407^{free}$ & $0.13$\tabularnewline
\hline \hline
$G_{17}$ & $1$& $-5426$ & $1077^{\pi N}$ & $1215^{\pi\pi N}$& $0.56$ & $G_{39}$ & $1$ &$869$  & $1077^{\pi N}$& $1688^{K \Sigma}$ &$0.57$ \tabularnewline
 &  $1$& $-1016$& $1077^{\pi N}$ & $1486^{\eta N}$  & $0.46$ & & $1$ & $-1701$ &  $1077^{\pi N}$ &$1517^{free}$  &$0.33$ \tabularnewline
 & $1$ & $-527 $ & $1077^{\pi N}$ &  $1462^{free}$& $0.34$ & & &  & &  & \tabularnewline
\hline \hline
$G_{19}$ & $1$& $-426$ & $1077^{\pi N}$ & $1486^{\eta N}$& $0.27$ & $H_{311}$ & $1$ &$-636$  & $1077^{\pi N}$& $1215^{\pi \pi N}$ &$0.64$ \tabularnewline
 &  $1$& $-404$& $1077^{\pi N}$ & $1611^{K\Lambda}$  & $0.26$ & & $1$ & $204$ &  $1077^{\pi N}$ &$1688^{K\Sigma}$  &$0.61$ \tabularnewline
 & $1$ & $-166$ & $1077^{\pi N}$ &  $1646^{free}$& $0.25$ & & $1$ & $-6762$ & $ 1077^{\pi N}$& $1828^{free}$ & $0.33$ \tabularnewline
\hline \hline
$H_{19}$ & $1$& $222$ & $1077^{\pi N}$ & $1486^{\eta N}$& $0.73$ &  &  & & &  & \tabularnewline
 &  $1$& $69$& $1077^{\pi N}$ & $1611^{K\Lambda}$  & $0.71$ & & & &  & & \tabularnewline
 & $1$ & $56$ & $1077^{\pi N}$ &  $1611^{free}$& $0.71$ & & &  & &  & \tabularnewline
\cline{1-6} \cline{1-6}
\end{tabular}
\end{table*}


\begin{table*}[ht]
\caption{\footnotesize \label{tab:pole2LPpar}  Parameters from L+P expansion are given for WI08 SE solutions. 
$N_r$ is number of resonance poles, $x_P, x_Q, x_R$ are branch points in $MeV$.}
\begin{tabular}{c|crclc|lcrclc}
\hline \hline
\multicolumn{12}{c}{Source}\tabularnewline
\multicolumn{12}{c}{WI08 SE L+P}\tabularnewline
\hline 
\hline 
PW & $N_{r}$ & $x_{P}$ & $x_{Q}$ & $x_{R}$ & $\chi^{2}_{dp}$ & PW & $N_{r}$ & $x_{P}$ & $x_{Q}$ & $x_{R}$ & $\chi^{2}_{dp}$\tabularnewline
\hline \hline
$S_{11}$ & $3$ & $-532$ & $1077^{\pi N}$ & $1215{}^{\pi\pi N}$ & $3.24$ & $S_{31}$ & $2$ & $405$ & $1077^{\pi N}$ & $1215^{\pi\pi N}$ & $2.76$\tabularnewline
 & $3$ & $-1.562$ & $1077^{\pi N}$ & $1486^{\eta N}$ & $3.21$ &  & $2$ & $-1498$ & $1077^{\pi N}$ & $1370^{Real(\pi\Delta)}$ & $2.73$\tabularnewline
 & $3$ & $-961$ & $1077^{\pi N}$ & $1498^{free}$ & $3.05$ &  & $2$ & $-524$ & $1077^{\pi N}$ & $1708^{Real(\rho N)}$ & $2.96$\tabularnewline
 &  &  &  &  &  &  & $2$ & $589$ & $1077^{\pi N}$ & $1167^{free}$ & $2.38$\tabularnewline
\hline 
\hline 
$P_{11}$ & $3$ & $-1369$ & $1077^{\pi N}$ & $1215^{\pi\pi N}$ & $2.61$ & $P_{31}$ & $1$ & $-67941$ & $1077^{\pi N}$ & $1215^{\pi\pi N}$ & $2.68$\tabularnewline
 & $3$ & $809$ & $1077^{\pi N}$ & $1370^{Real(\pi\Delta)}$ & $2.62$ &  & $1$ & $-27980$ & $1077^{\pi N}$ & $1214^{free}$ & $2.67$\tabularnewline
 & $3$ & $-2498$ & $1077^{\pi N}$ & $1202^{free}$ & $2.57$ &  &  &  &  &  & \tabularnewline
\hline 
\hline 
$P_{13}$ & $1$ & $-1798$ & $1077^{\pi N}$ & $1215^{\pi\pi N}$ & $1.59$ & $P_{33}$ & $2$ & $-11292$ & $1077^{\pi N}$ & $1215^{\pi\pi N}$ & $2.83$\tabularnewline
 & $1$ & $-68$ & $1077^{\pi N}$ & $1370^{Real(\pi\Delta)}$ & $1.68$ &  & $2$ & $-14313$ & $1077^{\pi N}$ & $1370^{Real(\pi\Delta)}$ & $2.85$\tabularnewline
 & $1$ & $-1806$ & $1077^{\pi N}$ & $1229^{free}$ & $1.58$ &  & $2$ & $-9413$ & $1077^{\pi N}$ & $1217^{free}$ & $2.78$\tabularnewline
\hline 
\hline
$D_{13}$ & $2$ & $-7313$ & $1077^{\pi N}$ & $1215^{\pi\pi N}$ & $2.86$ & $D_{33}$ & $1$ & $48$ & $1077^{\pi N}$ & $1215^{\pi\pi N}$ & $2.97$\tabularnewline
 & $2$ & $-510$ & $1077^{\pi N}$ & $1370^{Real(\pi\Delta)}$ & $3.17$ &  & $1$ & $-1335$ & $1077^{\pi N}$ & $1370^{Real(\pi\Delta)}$ & $3.17$\tabularnewline
 & $2$ & $-937$ & $1077^{\pi N}$ & $1708^{Real(\rho N)}$ & $3.14$ &  & $1$ & $-86102$ & $1077^{\pi N}$ & $1227^{free}$ & $2.59$\tabularnewline
 & $2$ & $-17258$ & $1077^{\pi N}$ & $1215^{free}$ & $2.81$ &  &  &  &  &  & \tabularnewline
\hline 
\hline 
$D_{15}$ & $1$ & $-326$ & $1077^{\pi N}$ & $1215^{\pi\pi N}$ & $2.46$ & $D_{35}$ & $1$ & $-59021$ & $1077^{\pi N}$ & $1688^{K\Sigma}$ & $4.48$\tabularnewline
 & $1$ & $-2159$ & $1077^{\pi N}$ & $1370^{Real(\pi\Delta)}$ & $2.53$ &  & $1$ & $-505$ & $1077^{\pi N}$ & $1679^{free}$ & $3.78$\tabularnewline
 & $1$ & $-295$ & $1077^{\pi N}$ & $1207^{free}$ & $2.44$ &  &  &  &  &  & \tabularnewline
\hline 
\hline 
$F_{15}$ & $2$ & $-2614$ & $1077^{\pi N}$ & $1215^{\pi\pi N}$ & $2.77$ & \multicolumn{1}{c}{$F_{35}$} & $1$ & $-6186$ & $1077^{\pi N}$ & $1215^{\pi\pi N}$ & $3.01$\tabularnewline
 & $2$ & $-2557$ & $1077^{\pi N}$ & $1370^{Real(\pi\Delta)}$ & $2.63$ &  & $1$ & $-10682$ & $1077^{\pi N}$ & $1708^{Real(\rho N)}$ & $3.04$\tabularnewline
 & $2$ & $-4442$ & $1077^{\pi N}$ & $1708^{Real(\rho N)}$ & $2.77$ &  & $1$ & $-3142$ & $1077^{\pi N}$ & $1360^{free}$ & $2.98$\tabularnewline
 & $2$ & $-1401$ & $1077^{\pi N}$ & $1231^{free}$ & $2.34$ &  &  &  &  &  & \tabularnewline
\hline 
\hline 
$F_{17}$ & $1$ & $-3903$ & $1077^{\pi N}$ & $1486^{\eta N}$ & $2.84$ & \multicolumn{1}{c}{$F_{37}$} & $1$ & $-4569$ & $1077^{\pi N}$ & $1370^{Real(\pi\Delta)}$ & $2.92$\tabularnewline
 & $1$ & $-2253$ & $1077^{\pi N}$ & $1611^{K\Lambda}$ & $2.95$ &  & $1$ & $-1999$ & $1077^{\pi N}$ & $1708^{Real(\rho N)}$ & $2.89$\tabularnewline
 & $1$ & $700$ & $1077^{\pi N}$ & $1669^{free}$ & $2.81$ &  & $1$ & $-2036$ & $1077^{\pi N}$ & $1384^{free}$ & $2.81$\tabularnewline
\hline \hline
$G_{17}$ & $1$ & $-179$ & $1077^{\pi N}$ & $1215^{\pi\pi N}$ & $2.24$ & \multicolumn{1}{c}{$G_{39}$} &  &  &  &  & \tabularnewline
 & $1$ & $-169$ & $1077^{\pi N}$ & $1486^{\eta N}$ & $1.75$ &  & &  &No results  & &\tabularnewline
 & $1$ & $18$ & $1077^{\pi N}$ & $1473^{free}$ & $1.52$ &  &  &  &  &  & \tabularnewline
\hline \hline
$G_{19}$ & $1$ & $-169$ & $1077^{\pi N}$ & $1486^{\eta N}$ & $2.26$ & \multicolumn{1}{c}{$H_{311}$} &  &  &  &  & \tabularnewline
 & $1$ & $-3367$ & $1077^{\pi N}$ & $1611^{K\Lambda}$ & $2.23$ &  & &  &No results  & &\tabularnewline
 & $1$ & $-4488$ & $1077^{\pi N}$ & $1620^{free}$ &  $2.21$&  &  &  &  &  & \tabularnewline
\hline \hline
$H_{19}$ & $1$ & $-1543$ & $1077^{\pi N}$ & $1486^{\eta N}$ & $2.15$ & \multicolumn{1}{c}{} &  &  &  &  & \tabularnewline
 & $1$ & $-445$ & $1077^{\pi N}$ & $1611^{K\Lambda}$ & $2.16$ &  & &  & & &\tabularnewline
 & $1$ & $-133$ & $1077^{\pi N}$ & $1513^{free}$ &  $2.11$&  &  &  &  &  & \tabularnewline
\cline{1-6}
\end{tabular}
\end{table*}


\clearpage
\begin{thebibliography}{9}

\bibitem{PDG} K.A. Olive et al. (Particle Data Group),
      Chin. Phys. C, \textbf{38}, 090001 (2014).     
\bibitem{Doering} M. D\"{o}ring, C. Hanhart, F. Huang, S. Krewald, and U.-G. Meissner, Nucl. Phys. \textbf{A829}, 170 (2009), and references therein.
\bibitem{EBAC} N. Suzuki,B. Juli\'{a}-D\'{i}az, H. Kamano, T.-S. H. Lee, A. Matsuyama, and T. Sato, Phys. Rev. Lett. \textbf{104}
042302 (2010); H. Kamano, T.-S. H. Lee, A. Matsuyama, T. Sato, N. Suzuki, Phys.Rev. \textbf{C80}, 025207 (2009), and references therein. 
\bibitem{CMB} R. E. Cutkosky, C. P. Forsyth, R. E. Hendrick, and R. L. Kelly, Phys. Rev. \textbf{D 20}, 2839 (1979).
\bibitem{Zagreb} M. Batini\'{c}, I. \v{S}laus, A. \v{S}varc, and B. M. K. Nefkens, Phys. Rev. \textbf{C 51}, 2310 (1995); M. Batini\'{c} et al., Phys. Scr. \textbf{58}, 15 (1998).
\bibitem{Bonn} A. V. Anisovich, R. Beck, E. Klempt, V. A. Nikonov, A. V. Sarantsev, U. Thoma,  Eur.Phys.J. \textbf{A48}, 15 (2012), and references therein.
\bibitem{Hoehler93} G. H\"{o}hler, $\pi$N Newsletter \textbf{9}, 1 (1993).
\bibitem{Kelkar} N.G. Kelkar, M. Nowakowski,  Phys.Rev. \textbf{A78}, 012709 (2008), and references therein. 
\bibitem{ChewMandelstam} G. F. Chew and S. Mandelstam, Phys. Rev. \textbf{119}, 467 (1960).
\bibitem{Ceci2008} S. Ceci, J. Stahov, A. \v{S}varc, S. Watson, and B. Zauner, Phys. Rev. \textbf{D 77}, 116007 (2008).
\bibitem{Masjuan2013}P. Masjuan , J. J. Sanz-Cillero,  Eur.Phys.\textbf{J. C73 }(2013) 2594.
\bibitem{GWUWI08} R. L. Workman, R. A. Arndt, W. J. Briscoe, M. W. Paris, and I. I. Strakovsky,  
Phys. Rev. \textbf{C 86}, 035202 (2012).
\bibitem{GWU2} http://gwdac.phys.gwu.edu/
\bibitem{GWU} R. A. Arndt, W. J. Briscoe, I. I. Strakovsky, and R. L. Workman, Phys. Rev. \textbf{C 74}, 045205 (2006). 
\bibitem{GWU1} R. A. Arndt,W. J. Briscoe, I. I. Strakovsky, R. L.Workman, and M. M. Pavan, Phys. Rev.\textbf{ C 69}, 035213 (2004).
\bibitem{LP1} A. \v{Svarc}, M. Had\v{z}imehmedovi\'{c}, H. Osmanovi\'{c}, J. Stahov, L. Tiator, and R. L. Workman, 
        Phys. Rev. \textbf{C 88}, 035206 (2013).
\bibitem{LP2}A. \v{Svarc}, M. Had\v{z}imehmedovi\'{c}, R. Omerovi\'{c}, H. Osmanovi\'{c}, and J. Stahov,
        Phys. Rev. \textbf{C 89}, 045205 (2014).
\bibitem{LP3} A. \v{Svarc}, M. Had\v{z}imehmedovi\'{c}, H. Osmanovi\'{c}, J. Stahov, L. Tiator, and R. L. Workman, Phys. Rev. \textbf{C 89}, 065208 (2014).
\bibitem{Mittag-Leffler}
        Michiel Hazewinkel: \emph{Encyclopaedia of Mathematics}, Vol.6,  Springer, 31. 8. 1990, pg.251.
\bibitem{Ciulli}
        S. Ciulli and J. Fischer, Nucl. Phys. 24, 465 (1961).
\bibitem{Ciulli1}
        I. Ciulli, S. Ciulli, and J. Fischer, Nuovo Cimento \textbf{23}, 1129 (1962).
\bibitem{Pietarinen} E. Pietarinen, Nuovo Cimento Soc. Ital. Fis. \textbf{12A}, 522
(1972).
\bibitem{Hoehler84}
        G. H\"{o}hler, \emph{Pion Nucleon Scattering}, Part 2, Landolt-B\"ornstein:
        Elastic and Charge Exchange Scattering of Elementary Particles, Vol. 9b (Springer-Verlag, Berlin, 1983).
\bibitem{ConMap-Use} C. G. Boyd, B. Grinstein, and R. F. Lebed,  Phys. Rev. Lett. \textbf{74}, 4603 (1995); R. J. Hill, and G. Paz, Phys.
Rev. \textbf{D 82}, 113005 (2010).
\bibitem{Gribov} V.N. Gribov, 'Strong Interactions of hadrons at high energies', Cambridge U. Press, 2009.
\bibitem{Ceci2011} S. Ceci, M. D\"{o}ring, C. Hanhart, S. Krewald, U.-G. Mei\ss{}ner, and A. \v{S}varc, Phys. Rev.  \textbf{C 84},
015205 (2011).
\bibitem{Roenchen2013} D. R\"{o}nchen, M. D\"{o}ring, F. Huang, H. Haberzettl, J. Haidenbauer, C. Hanhart, S. Krewald, U.-G. Meissner, and
K. Nakayama, Eur. Phys. J. \textbf{A 49}, 44 (2013).
\bibitem{cmb} R.L. Kelly and R.E. Cutkosky, Phys. Rev. \textbf{D20}, 2782 (1979);
R.E. Cutkosky, R.E. Hendricks, J.W. Alcock, Y.A. Chao, R.G. Lipes, J.C. Sandusky, and R.L. Kelly,
Phys. Rev. \textbf{D20}, 2804 (1979); R.E. Cutkosky, C.P. Forsyth, R.L. Hendrick, and R.L. Kelly,
Phys. Rev. \textbf{D20}, 2839 (1979).


\end{thebibliography}
\end{document}